\documentstyle[mprocl]{article}

\def\be{\begin{equation}}
\def\ee{\end{equation}}
\def\bea{\begin{eqnarray}}
\def\eea{\end{eqnarray}}

\begin{document}

\title{Cylindrical Domain Walls and Gravitational Waves\\
  \smallskip
  {\normalsize --Einstein Rosen wave emission from momentarily static
    initial configuration --}}

\author{Kouji NAKAMURA${}^{*}$ and Hideki ISHIHARA${}^{**}$}

\address{${}^{*}$  Advanced Science Research Center,
  Japan Atomic Energy Research Institute,\\
  Tokai, Naka, Ibaraki 319-11, Japan}

\address{${}^{**}$
  Department of Physics,
  Tokyo Institute of Technology, \\
  Oh-Okayama Meguro-ku, Tokyo 152, Japan
  }


\maketitle

\abstracts{
A self-gravitating cylindrical domain wall is considered as an example
of non-spherical wall to clarify the interaction between a domain wall
and gravitational waves. We consider the time evolution from a
momentarily static initial configuration within an infinitesimal time
interval using the metric junction formalism. We found that the wall
with a large initial radius radiates large amplitude of the
gravitational waves and undergoes its large back reaction. 
}

\section{Introduction}

Domain walls could be formed as a topological defect associated with
the discrete symmetry breaking in the universe. It is thought that, in
the treatment neglecting gravity, an oscillatory motion of the wall
becomes to be a source of gravitational waves even in the first order
with respect to its oscillation amplitude. If so, the energy loss rate
can be estimated by the quadrupole formula for gravitational wave
emission\cite{VEV}, since it seems that the radiation reaction can be
negligible.

However, the recent investigations\cite{KoIshiFuji} show that there is 
no dynamical freedom in the domain wall coupled to gravitational
field. In Ref.2, a spherically symmetric self-gravitating domain wall
is considered as a background and the general relativistic
perturbations are studied. It is found that the domain wall does not
emit gravitational waves spontaneously by its free
oscillation. However, it is not clear whether the result is caused by
the spherical symmetry or not. Indeed, there is no local mode of
gravitational waves in spherically symmetric spacetime which is used
as a background in the soluble models.

In this talk, we consider the interaction between a domain wall and
gravitational waves in a less symmetric model. We consider a
self-gravitating thin wall with cylindrical symmetry as an example of
non-spherical one. We use Israel's metric junction
formalism\cite{Israel} to analyze the self-gravitating thin wall. So,
one can easily see that the motion of wall directly leads to the
gravitational wave emission.

\section{Cylindrical Self-gravitating Domain Wall}

Now, we consider the self-gravitating cylindrical domain wall. For
simplicity, we assume that spacetime consists of two vacuum regions
with cylindrical symmetry. A cylindrically symmetric spacetime is
described by Weyl canonical form
\begin{equation}
  \label{Weyl-cano}
  ds^{2} = e^{2(\gamma - \psi)} (- dt^{2} + dr^{2}) + e^{2\psi} dz^{2}
   + r^{2} e^{-2\psi}d\phi^{2}.
\end{equation}
From this metric, the Einstein equations can be reduced to
\begin{eqnarray}
  \label{ER-wave}
  && \partial_{t}^{2}\psi - \frac{1}{r}\partial_{r}\left(r\partial_{r}
  \psi\right) = 0, \\
  \label{gammar}
  && \partial_{r}\gamma = r \left((\partial_{t}\psi)^{2} +
  (\partial_{r}\psi)^{2}\right), \\
  \label{gammat}
  && \partial_{t}\gamma = 2 r (\partial_{t}\psi) (\partial_{r}\psi).
\end{eqnarray}
The wave solution $\psi$, which is well-known as Einstein Rosen wave
(ERW), is regarded as a cylindrically symmetric mode of gravitational
wave. On the other hand, $\gamma$ corresponds to the gravitational
potential of a cylindrically symmetric spacetime.

The junction conditions for the wall with the radius $R$ are given by 
\begin{eqnarray}
  \label{junction-tautau}
  && \left(\ddot{R} - \dot{R}\dot{\psi} + R \dot{\psi}^2\right)
  \left(\frac{1}{X_{+}} - \frac{1}{X_{-}}\right) +
    R\left(\frac{\left(D_{\perp}\psi\right)_{+}^2}{X_{+}} -
    \frac{\left(D_{\perp}\psi\right)_{-}^2}{X_{-}}\right) = -
  2\lambda, \\
  \label{junction-psi}
  && (D_{\perp}\psi)_{+} - (D_{\perp}\psi)_{-} = - \lambda, \\
  \label{junction-r}
  && X_{+} - X_{-} = - 2 \lambda R
\end{eqnarray}
where $X_{+} = \sqrt{\dot{R}^{2} + e^{-2\gamma_{+}}}$ and $X_{-} =
\sqrt{\dot{R}^{2} + 1}$. In
(\ref{junction-tautau})-(\ref{junction-r}), $(D_{\perp}\psi)_{+}$ 
($(D_{\perp}\psi)_{-}$) is the derivative of $\psi$ just outside
(inside) the wall along the normal direction to the world hyper sheet
of the wall, $\lambda$ is the wall energy density, the dot denotes the
derivative with respect to the proper time $\tau$ on the wall, and
$\gamma_{+}$ corresponds to the deficit angle just outside the wall.
The condition (\ref{junction-psi}) is the boundary condition for the
ERW on the wall, (\ref{junction-r}) is the equation of the wall
motion. Equation (\ref{junction-tautau}) is the first derivative of
(\ref{junction-r}) with respect to $\tau$.

As the simplest case, we consider the momentarily static initial
configuration, {\it i.e.}, $\dot{R} =0,~ \partial_{t}\psi =
\partial_{t}\gamma = 0$, with the initial wall radius $R_{i}$. From
the regularity at the symmetric axis, the inside region of the wall
should be momentarily Minkowski spacetime, {\it i.e.},
\begin{equation}
  \label{initial_static_configuration_int}
  \psi_{{\cal M}_{-}} = 0, \quad
  \gamma_{{\cal M}_{-}} = 0 ,
\end{equation}
while from (\ref{junction-psi}), outside region is described by
\begin{equation}
  \label{initial_static_configuration}
  \psi_{{\cal M}_{+}} 
                = - \kappa_{+}\ln\left(\frac{r}{R_{i}}\right),  \;\;\;\;\;\;
  \gamma_{{\cal M}_{+}} 
                = \gamma_{+} + \kappa^{2}_{+} \ln\left(\frac{r}{R_{i}}\right),
\end{equation}
where $\kappa_{+} = \lambda R_{i}/(1 - 2\lambda R_{i})$, $\gamma_{+} =
- \ln (1 - 2 \lambda R_{i})$, and the variables with the suffix ${\cal
  M}_{\pm}$ mean the values in the outside and inside regions of the
wall, respectively. From (\ref{junction-tautau}), the initial
configuration (\ref{initial_static_configuration_int}) and
(\ref{initial_static_configuration}) give us the initial acceleration
$\ddot{R}_{i} = - (2 - 3 \lambda R_{i})/R_{i}$. The time evolution is
given by solving the Einstein equations (\ref{ER-wave})-(\ref{gammat})
and junction conditions (\ref{junction-tautau})-(\ref{junction-r}).
The solution within the infinitesimal time interval is given in the
form\cite{kouchan}  
\begin{eqnarray}
  \label{Taylor-series-plus}
  \psi_{{\cal M}_{+}} &=& - \kappa_{+}\ln\frac{r}{R_{i}} +
  \frac{1}{2}\left(\frac{B}{r^{1/2}}\right) (\Delta u)^{2} +
  O((\Delta u)^3),  \\
  \label{Taylor-series-minus}
  \psi_{{\cal M}_{-}} &=& - \frac{1}{2}\left(\frac{B}{r^{1/2}}\right)
  (\Delta v)^{2} + O((\Delta v)^3), \\
  \label{r-solution}
  R(\tau) &=& R_{i} + \frac{1}{2} \ddot{R}_{i} (\Delta\tau)^{2} +
  \frac{1}{3} \frac{(1 - 2 \lambda R_{i})^{3}\kappa_{+}B}{\lambda
    R_{i}\sqrt{R_{i}}} (\Delta\tau)^{3} + O((\Delta\tau)^{4}).
\end{eqnarray}
where  $\Delta u$ ($\Delta v$) is the infinitesimal retarded
(advanced) time interval from the initial time and the initial
amplitude of ERW, $B$, is given by 
\begin{equation}
  \label{amplitude}
        B = - \lambda \ddot{R}_{i}/
                \left(2(1-2 \lambda R_{i})^{3}\sqrt{R_{i}}\right). 
\end{equation}
We can see the later behavior of the solution by solving
(\ref{ER-wave})-(\ref{junction-r}) order by order.

From the solution (\ref{Taylor-series-plus})-(\ref{amplitude}), we
can easily see that the initial amplitude $B$ of ERW is determined by
the initial acceleration $\ddot{R}_{i}$ of the wall and, on the other
hand, the motion of the wall is directly affected by the amplitude
$B$. Equation (\ref{r-solution}) shows that the effect of back
reaction appears in the order of $(\Delta\tau)^{3}$. Since
$\ddot{R}_{i}$ is negative initially and the coefficient of the order
of $(\Delta\tau)^{3}$ in (\ref{r-solution}) is positive, the back
reaction by wave emission reduces the wall's acceleration.
Furthermore, we can see that if $R_{i} \ll 1/(2\lambda)$, the
amplitude of ERW is small and its back reaction is small. Indeed, the
back reaction does not become important until the wall collapse to
zero radius in this case. However, if the initial radius becomes large
as $R_{i} \rightarrow 1/(2\lambda )$, ERW is emitted with large
amplitude and the back reaction is not negligible.

\section{Summary}

We have investigated the behavior of a self-gravitating cylindrical
domain wall. Originally, the problem to solve this system corresponds
to the radiation reaction problem and we must not neglect the emission
of gravitational waves. Here, we have solved the time evolution of the 
system within an infinitesimal time interval from the momentarily
static initial configuration using the metric junction formalism.
First, we have found that the acceleration of the wall decreases by
the back reaction via the emission of gravitational waves. Secondly,
the wall with small initial radius emits gravitational waves with
small amplitude and then its back reaction can be negligible. On the
other hand, we have seen that the wall with large initial radius emits
gravitational waves with large amplitude and the motion of the wall is
altered considerably by its back reaction. Thus, it is important to
take the interaction between the wall and gravitational waves into
account. 


\end{document}